\documentclass[a4paper,fleqn,usenatbib]{mnras}

\usepackage{newtxtext,newtxmath}

\usepackage{ae,aecompl}

\usepackage{graphicx}	
\usepackage{amsmath}	
\usepackage{amssymb}	
\newcommand{\LL}{\mathcal{L}}

\newcommand{\hMpc}{ h^{-1}{\rm Mpc}}
\newcommand{\ihMpc}{ h\,{\rm Mpc}^{-1}}

\title[Likelihood Emulation]{Cosmological parameter estimation via iterative emulation of likelihoods}

\author[]{Marcos Pellejero-Iba\~nez$^{1}$\thanks{E-mail: mpellejero@dipc.org}, Raul E. Angulo$^{1,2}$\thanks{E-mail: reangulo@dipc.org}, Giovanni Aric\'o$^{1}$\thanks{E-mail: giovanni$\_$arico001@ehu.es}, Matteo Zennaro$^{1}$, 
\and Sergio Contreras$^{1}$ \& Jens St\"ucker$^{1}$.
\\
$^{1}$Donostia International Physics Center (DIPC), Paseo Manuel de Lardizabal, 4, 20018 Donostia-San Sebasti\'an, Spain\\
$^{2}$IKERBASQUE, Basque Foundation for Science, 48013, Bilbao, Spain.}

\date{Accepted XXX. Received YYY; in original form ZZZ}

\pubyear{2019}

\begin{document}
\label{firstpage}
\pagerange{\pageref{firstpage}--\pageref{lastpage}}
\maketitle

\begin{abstract}
The interpretation of cosmological observables requires the use of increasingly sophisticated theoretical models. Since these models are becoming computationally very expensive and display non-trivial uncertainties, the use of standard Bayesian algorithms for cosmological inferences, such as MCMC, might become inadequate. Here, we propose a new approach to parameter estimation based on an iterative Gaussian emulation of the target likelihood function. This requires a minimal number of likelihood evaluations and naturally accommodates for stochasticity in theoretical models. We apply the algorithm to estimate 9 parameters from the monopole and quadrupole of a mock power spectrum in redshift space. We obtain accurate posterior distribution functions with approximately 100 times fewer likelihood evaluations than an affine invariant MCMC, roughly independently from the dimensionality of the problem. We anticipate that our parameter estimation algorithm will accelerate the adoption of more accurate theoretical models in data analysis, enabling more comprehensive exploitation of cosmological observables.
\end{abstract}

\begin{keywords}
Cosmology -- large-scale structure of Universe -- cosmological parameters
\end{keywords}



\section{Introduction}

Estimation of cosmological parameters is an essential part of modern cosmology. Constraints on the main background parameters of the standard model have provided the most significant evidence for dark matter and the accelerated expansion of the universe (see e.g. \citealt{Riess:1998cb}, \citealt{planck2015-xiii} and \citealt{Alam:2016hwk}). Furhermore, tensions between parameters can even be used as indications of yet-unknown physics \citep[e.g.][]{Aghanim:2018eyx,Joudaki:2019pmv,Sola:2016ecz}.

In cosmological data analysis, parameters are usually estimated using Bayesian statistics \citep{Trotta:2008qt}. Widely-used algorithms are, e.g, Monte-Carlo Markov-Chains (see \citealt{Christensen:2001gj}, or \citealt{doi:10.1002/0470011815.b2a14021} for a review), Population Monte Carlo \citep{PopMonteCarlo,Wraith2009}, Nested sampling (\citealt{MR2282208}), VEGAS (\citealt{PETERLEPAGE1978192}), Approximate Bayesian Computation \citep{Ishida2015,Akeret2015}, Density estimation likelihood-free inference \citep{Fan2013,Papamakarios2016}, Hamiltonian Monte Carlo \citep{DUANE1987216}, and multiple variants have been proposed \citep[e.g.][]{Foreman2013, BAMBI,goodman2010,Elson:2006ui,Calvo2005,Kitaura2013,Hernandez-Sanchez2019}. Although they vary in the details, all these algorithms rely on a large number of evaluations of a theoretical model for the particular summary statistic of interest. This, unfortunately, might pose a difficult challenge for future data analyses.

We have entered an era in the study of cosmology in which the precision of the measurements forces the development and use of increasingly sophisticated theoretical models. This is particularly true for large-scale structure (LSS) analysis. In this area, models aim at predicting the non linear evolution of the cosmological density and velocity fields, as well as the distribution of biased tracers such as galaxies or quasars. To model the underlying physics accurately they further include  gravitational interactions and galaxy formation physics among other baryonic effects. Although the nature of the modelling might vary wildly -- from perturbation theory to $N$-body simulations --, they are all becoming computationally expensive to evaluate and have non-negligible uncertainties that may vary with cosmological parameters. This will likely make traditional techniques for parameter estimation inadequate for future LSS analyses. 

Firstly, the expensiveness of model evaluations together with a large number of evaluations required for convergence (due to a large hyper-parameter space) might make the problem computationally unfeasible. Secondly, as we will later see in this work, noise in the theory-model translates into a noisy posterior function with multiple local maxima in the likelihood, which heavily slows down convergence in MCMC methods. Examples of the former are high dimensional integrals in perturbation theory or huge number of force evaluations in simulations, and for the latter, the impact of higher order contributions in perturbation theory or cosmic variance in simulations.

A possible way to tackle these problems is with the aid of less accurate models and posterior distributions \citep{Taruya:2012ut,Chuang:2016uuz,Pellejero-Ibanez:2016ypj,Moews:2019vlt}, or by emulating a specific summary statistic. Emulation is a form of interpolation where a small number of (expensive) function evaluations are used to predict values throughout a given parameter space. In this way, model evaluation can become very fast and traditional Bayesian samplers can be used. Recently, this technique has gained a lot of attention in the LSS community; it has been used in modelling the small-scale matter power spectrum \citep[see e.g.][]{Heitmann:2009cu,Knabenhans2019,Giblin2019}, the galaxy power spectrum and correlation function \citep{Kwan:2013jva,Zhai:2018plk}, weak lensing peak counts and power spectra \citep{Liu:2014fzc,Petri:2015ura,Manrique-Yus:2019hqc,Giblin2018}, the 21 cm power spectrum \citep{Jennings:2018eko} and the halo mass function \citep{McClintock:2018uyf}. On the other hand, emulators are notoriously difficult to build in high dimensions, and have an uncertainty structure that might propagate to cosmological parameter space. Although the uncertainty can be decreased in certain parts of the parameter space iteratively \citep{Rogers2019}, the emulator approach would need to change (or at least be validated) every time a new summary statistic or range of scales is considered.

In this work we propose an alternative strategy for estimating parameters  that circumvents the aforementioned problems. Our approach employs a Gaussian emulation directly on the target likelihood function while naturally accounting for the uncertainty in the theory model. This emulation is initially built with a small number of model evaluations and then its faithfulness and accuracy are iteratively increased by focusing new evaluations on relevant regions of the parameter space that also display large uncertainty. Similar ideas were already proposed in \cite{Leclercq2018} (based on the works by \citealt{Gutmann2015} and \citealt{Jarvenpaa2017}) for the Likelihood-free inference case and by \cite{Velden2019} (based on the works by \citealt{Goldstein2015} and \citealt{Vernon2014}) for the Bayes linear approach on an MCMC.

Emulation of likelihoods have been previously attempted. Specifically, \cite{McClintock2019} showed that if the high-dimensional Planck likelihood (\citealt{planck2015-xiii}) is known, then it can be accurately emulated with a relatively modest number of evaluations, whereas \cite{Aslanyan:2015} showed that iteratively emulating CMB likelihoods can yield up to factors of 3-5 less model evaluations than an MCMC. 

Here we build on those works by showing that an iterative Gaussian emulation can be successfully used for constraining parameters with $\sim 100$ times less model evaluations than a standard MCMC. Specifically, we are able to explore the constraints of the monopole and quadrupole of the galaxy power spectrum in 9 dimensions with approximately less than $700$ model evaluations.

This paper is organized as follows. In \S\ref{Sec:emulator} we discuss the main ideas in our proposed approach: Gaussian emulation of likelihoods, the connection with uncertain theoretical models in parameter estimation, and adaptively improving the accuracy of the emulation. In \S\ref{Sec:algorithm} we specify details of the algorithm we employ. In \S\ref{Sec:sampling} we apply the iterative emulator to constrain the free parameters of a model for the multipoles of the galaxy power spectrum in redshift space. In this section we also compare its performance against a standard MCMC approach, and discuss the impact of several choices in our algorithm. Finally, we summarise and conclude in \S\ref{Sec:Conclusions}. 

\section{Iterative Gaussian emulation of likelihoods}\label{Sec:emulator}

According to Bayes theorem, the posterior probability, $p(\boldsymbol{\theta}|\boldsymbol{d})$, of a set of parameters $\{\boldsymbol{\theta}\}$ given an observable, $\textbf{d}$, is 

 \begin{equation}
 p(\boldsymbol{\theta}|\boldsymbol{d}) = \LL(\textbf{d}|\boldsymbol{\theta})\,p(\boldsymbol{\theta}) / p(\boldsymbol{d}) 
 \end{equation}

\noindent where $p(\boldsymbol{\theta})$ is referred to as prior distribution, $\mathcal{L}$ is the likelihood and $p(\boldsymbol{d})$ is a normalisation called Bayesian evidence. If data uncertainties are described by a multivariate Gaussian, then $\LL$ takes the form: 

\begin{equation}
    \LL(\boldsymbol{d}|\boldsymbol{\theta}) \equiv\frac{1}{\sqrt{(2\pi)^n|C_d|}}\exp{\left(-\frac{1}{2}\chi^2(\boldsymbol{\theta})\right)}\, ,
\label{eq:likelihood}
\end{equation}

\noindent where $\chi^2 \equiv X(\boldsymbol{\theta})^\intercal C_d^{-1}X(\boldsymbol{\theta})$ and $X \equiv \textbf{d}-\textbf{t}(\boldsymbol{\theta})$, $\boldsymbol{t}$ are the predictions for $\textbf{d}$ given the parameters $\{\boldsymbol{\theta}\}$, and $C_d$ is the data covariance matrix. Hereafter, we will drop the explicit dependence on $\boldsymbol{\theta}$ unless explicitly needed.

The problem of parameter estimation can be summarised as finding the maximum of $\LL$ and computing posterior distribution functions. For this, traditional Bayesian algorithms typically require a large number of evaluations of $\textbf{t}$ at different sets of $\{\boldsymbol{\theta}\}$.

Depending on the characteristics of the theory $\textbf{t}$, these evaluations can be computationally heavy, thus making such analysis unfeasible. Additionally, $\textbf{t}$ might have a degree of stochasticity and biases in its predictions, both of which, in general, depend on $\{\boldsymbol{\theta}\}$.

Therefore, being able to reconstruct the likelihood from as few evaluations as possible and incorporating theory uncertainties is of crucial importance, and is the main goal of this paper.

In this section we outline the main ideas behind our approach. We start by recapping Gaussian emulation in \S\ref{sec:gp}, then we show how theory errors can be included in \S\ref{sec:noise}, and in \S\ref{sec:iter} discuss an iterative scheme to progressively improve the accuracy of our method.

\subsection{Gaussian Processes}\label{sec:gp}

Let us assume there is a set of values $\{\boldsymbol{x}\}$ for which we have evaluated the function $\boldsymbol{f} \equiv \log \LL(\boldsymbol{x})$, and another set, $\{ \boldsymbol{y}\}$, for which we would like to estimate the values $\boldsymbol{f}_* \approx \log \LL(\boldsymbol{y}) $ without evaluating the likelihood at those points. The problem can be solved by computing the probability distribution of the surrogate function $\boldsymbol{f}_*$, $p(\boldsymbol{f}_*|\boldsymbol{y},\boldsymbol{x},\boldsymbol{f})$. Gaussian emulation adopts the ansatz that $\boldsymbol{f}$ and $\boldsymbol{f}_*$  are jointly distributed Gaussian variates:

\begin{equation}
 p(\boldsymbol{f}_*,\boldsymbol{f}) \sim \mathcal{N} \left(
\begin{pmatrix}
    \boldsymbol{\mu}\\
    \boldsymbol{\mu}_*
\end{pmatrix},
\begin{pmatrix}
    K & K_*\\
    K^\intercal_* & K_{**}
\end{pmatrix} \right) \, .
\end{equation}

Here, $K$ refers to the the covariance of $\boldsymbol{f}$, $K_{**}$ to that of $\boldsymbol{f}_*$, and $K_{*}$ denotes the cross-covariance between $\boldsymbol{f}$ and $\boldsymbol{f}_*$. 

Thanks to the multivariate Gaussian theorem, we can use the joint distribution $p(\boldsymbol{f},\boldsymbol{f}_*)$ to compute the conditional distribution $p(\boldsymbol{f}_*|\boldsymbol{f})$, which also follows a Gaussian distribution with mean and covariance:

\begin{equation}
 \begin{cases}
\boldsymbol{\mu}_{\rm{cond}}=\boldsymbol{\mu} + K_*K^{-1}_{**}(f-\boldsymbol{\mu}_*) \\
\Sigma_{\rm{cond}} = K- K_*K^{-1}_{**}K^\intercal_*
\end{cases}\, .
\end{equation}

In Gaussian emulations, it is customary to assume $\boldsymbol{\mu}=\boldsymbol{\mu}_*=0$ since the flexibility provided by the covariances is enough to model $\boldsymbol{f}_*$ arbitrarily well. It is also common to assume that the form of the covariances is the same. A freedom in the problem is the functional form of the kernel function, or covariance, $K$. Throughout this work we will adopt a squared exponential kernel:

\begin{equation}
 K_{i,j}=\sigma^2_{\rm{k}} \exp \left[ -\frac{(x_i-x_j)^2}{2\,l^2_{\rm K}} \right] + \delta_D(i-j)\sigma^2_x\, ,
 \label{eq:kernel}
\end{equation}

\noindent where similarity is a monotonic function of the distance. The length parameter $l_{\rm K}$ controls the smoothness of the function and $\sigma^2_{\rm k}$ its amplitude. 

These parameters are both free but can be found by maximizing the marginal log-likelihood of the Gaussian process (see \citealt{Rasmussen:2005:GPM:1162254}):

\begin{equation}
 \log p(\boldsymbol{f}|x,K) = -\frac{1}{2}\boldsymbol{f}^T\,K^{-1}\,\boldsymbol{f}-\frac{1}{2}\log|K|-\frac{N}{2}\log(2\pi) \, ,
 \label{likelihood_gp}
\end{equation}

\noindent where $N$ is the number of known function evaluations (i.e. size of the $f$ vector). As a result, we obtain an estimate for the probability distribution of the surrogate function $f_*$, from which we can evaluate the most likely outcome of the emulated likelihood, $\LL_{\rm{emu}}$, and also its expected uncertainty, $\sigma[\LL_{\rm{emu}}]$. 

In principle, there is an extra parameter to consider, $\sigma_{x}$, the amount of noise you expect in the training evaluations $f$. It is also possible to include $\sigma_x$ in the maximization of Eq.~\ref{likelihood_gp}. However, in the next subsection we will show that this parameter is related to the uncertainty in the theory model, thus it can be set in advance. 

\subsection{$\chi^2$ distribution for uncertain model predictions}\label{sec:noise}

Let us consider a case in which the theoretical predictions for a given observable are not exact but have intrinsic uncertainties. This is likely the case for state-of-the-art models of the large-scale distribution of galaxies and matter in the Universe.

Specifically, for the case of models built from $N$-body simulations, uncertainties in the initial conditions (cosmic variance), arbitrariness in group finders, errors introduced by the finite accuracy of force calculation and time integration, approximations in the physics (e.g. AGN/SNe feedback), etc, in general will result in stochastic and biased predictions. Additional sources of errors arise from cosmology rescaling methods (e.g. \citealt{Angulo:2010}, Contreras et. al. in prep), emulators for LSS (e.g. \citealt{Kwan:2013jva,Zhai:2018plk}), or approximated gravity solvers (e.g. \citealt{2016MNRAS.463.2273F}). Similarly, in the case of perturbation theory, uncertainty can arise from the contribution of neglected orders, approximations in the equations of motions, and also from neglected physics such as galaxy formation (e.g. \citealt{Baldauf:2016sjb}).

For simplicity, we consider a case where the theoretical predictions, $\boldsymbol{t}(\boldsymbol{\theta})$, have correlated Gaussian noise given by $C_t$ and are biased by $\boldsymbol{\mu}$. In such cases, the covariance matrix needs to be modified so that the likelihood properly accounts for the uncertainty in the models $C \rightarrow C_d + C_t + \boldsymbol{\mu}^2$  \citep{Baldauf:2016sjb,Audren:2012vy,Sprenger:2018tdb}. We now discuss how the noise in $\boldsymbol{t}(\boldsymbol{\theta})$ propagates to the likelihood itself, which consequently affects its emulation.

We compute the expected distribution of $\chi^2(\boldsymbol{\theta})$ for many realizations of the theory. Following \cite{CIS-6533}, we start by defining the variable $Z$: 

\begin{equation}
    Z \equiv C_t\,C^{-1/2}\,X - C_t^{-1/2}\mu
\end{equation}

\noindent which is, thus, Normal distributed with identity variance and a zero mean, $ Z \sim \mathcal{N}(0,\mathbb{I})$. Under this change of variables, $\chi^2(\boldsymbol{\theta})$ can be written as:

\begin{equation}
    \chi^2(\theta) = \left( Z+C_t^{-1/2}\mu\right)^\intercal C_t^{1/2} C^{-1}C_t^{1/2} \left( Z + C_t^{-1/2}\mu\right).
\end{equation}

Invoking the spectral theorem, we perform a diagonalization $R^\intercal \left[ C_t^{1/2} C^{-1} C_t^{1/2}\right] R = {\rm{diag}}(\lambda_1,\cdots,\lambda_n)$, where $n$ is the size of the data vector, the $\lambda$s are the eigenvalues of $C_t C^{-1}$, and $R$ is the matrix of its eigenvectors. With this rotation in mind, we perform a last change of variables $U\equiv R^\intercal Z$. Note here that $U$ is still standard normally distributed since $R^\intercal R=\mathbb{I}$. Then

\begin{eqnarray}
    X^\intercal C X  &=& (U+\boldsymbol{b})^\intercal{\rm{diag}}(\lambda_1,\cdots,\lambda_n)(U+\boldsymbol{b}) \\ &=& \sum_{i=0}^{n} \lambda_i(U_i+b_i)^2 ,
\end{eqnarray}

\noindent where $\boldsymbol{b} \equiv (R^\intercal C_t^{-1/2}\boldsymbol{\mu})^\intercal$. In general, each of the summands follows a \emph{Gamma} distribution, since they are squared normally distributed variables with mean $b_i$ and standard deviation $\lambda_i$. However, the particular expression above can be written as a sum  of $n$ Gamma distributions, $\Gamma(1/2,2\lambda_i)$, plus the sum of the Normal distributions $\mathcal{N}(\lambda_ib_i^2,\lambda_i^2b_i^2)$. 

In general, the sum of Gamma functions does not have a closed form.\footnote{An exact solution for the general case in terms of recursive relations was found by \cite{mathai1982} and \cite{Moschopoulos1985}.} However, it can be approximated as a single Gamma distribution, $\sum_i\Gamma(1/2,2\lambda_i) \simeq \Gamma(k,\theta)$, where $k=(\sum_i\lambda_i)^2/\sum_i2\lambda_i^2$ and $\theta=\sum_i\lambda_i/k$ \citep{10.2307/3002019,10.2307/2332510}. 

A Gamma distribution approaches a Normal distribution for large shape parameter $k$. Thus, as long as we have enough degrees of freedom (large $n$) and none of them has a value of $\lambda_i$ much larger than the sum of the others ($\lambda_{i\rm{th}}\gg\sum_j\lambda_j$), this Gamma function can be well approximated by a Normal distribution $\mathcal{N}(\sum_i\lambda_i,\sum_i2\lambda_i^2)$. 

The above implies that $\chi^2(\theta)$ itself is approximately Normal distributed. The  expectation value and variance of the change in $\chi^2$ induced by the noise and biases
in a theory model are: 

\begin{equation}
E[\Delta \chi^2]= \sum_i \lambda_i(1+b_i^2),
\label{equation:mean}
\end{equation}

\begin{equation}
    {\rm{Var}}[\Delta \chi^2]= \sum_i \lambda_i^2(2+b_i^2).
\label{equation:variance}
\end{equation}

Note that in the case of deterministic models, $C_t = 0$ and $\lambda_i = 0$, thus the values of $\chi^2$ are unchanged. In the case of unbiased models, 
$\boldsymbol{\mu} = 0$ and $\boldsymbol{b}=0$, the best fit values will in general shift, but we expect this bias to be statistically insignificant. 

In summary, uncertain data models induce a Gaussian noise in the logarithm of the likelihood. This noise in $\chi^2(\boldsymbol{\theta})$ is given by Eq.~\ref{equation:variance} and can be incorporated within our Gaussian emulation by setting the magnitude of $\sigma_x$ in Eq.~\ref{eq:kernel}.

\begin{figure*}
\centering
\includegraphics[width=0.15\textwidth]{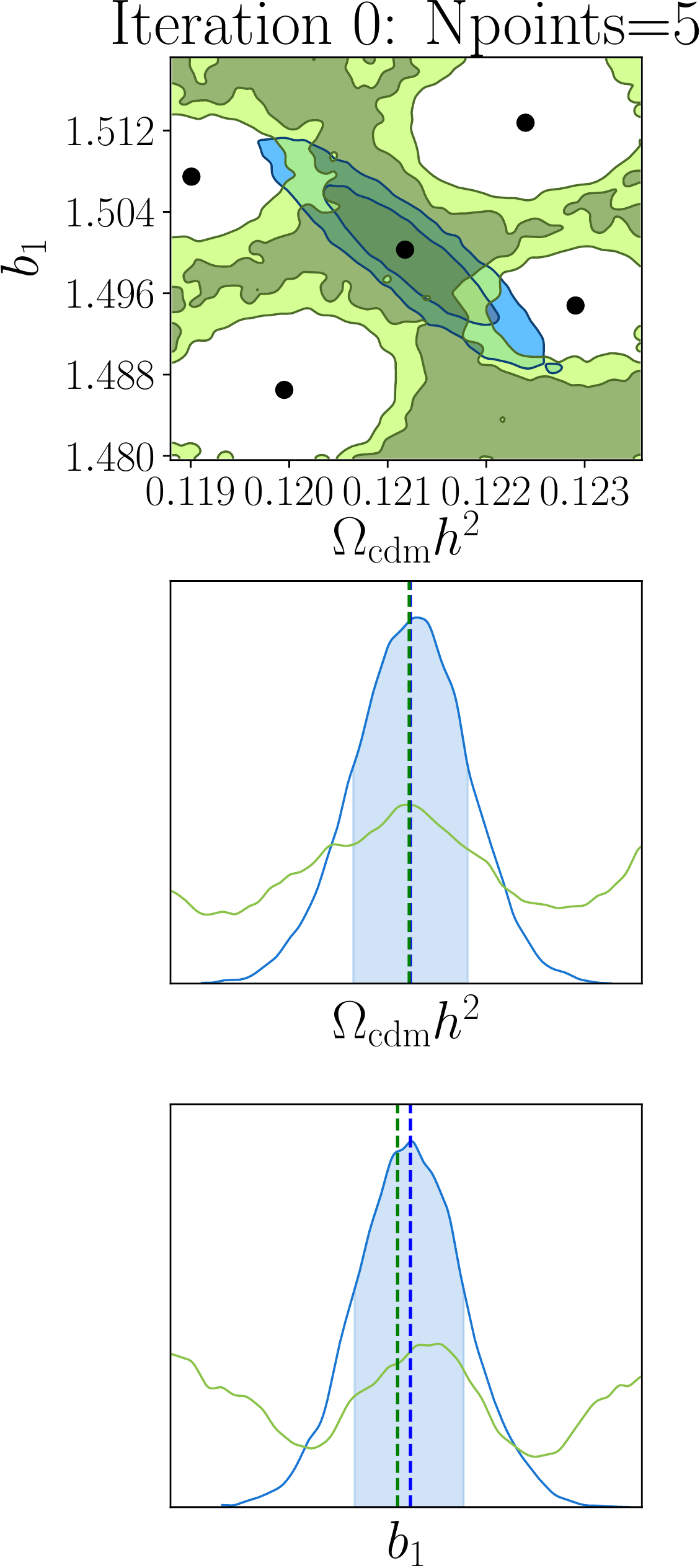}
\includegraphics[width=0.15\textwidth]{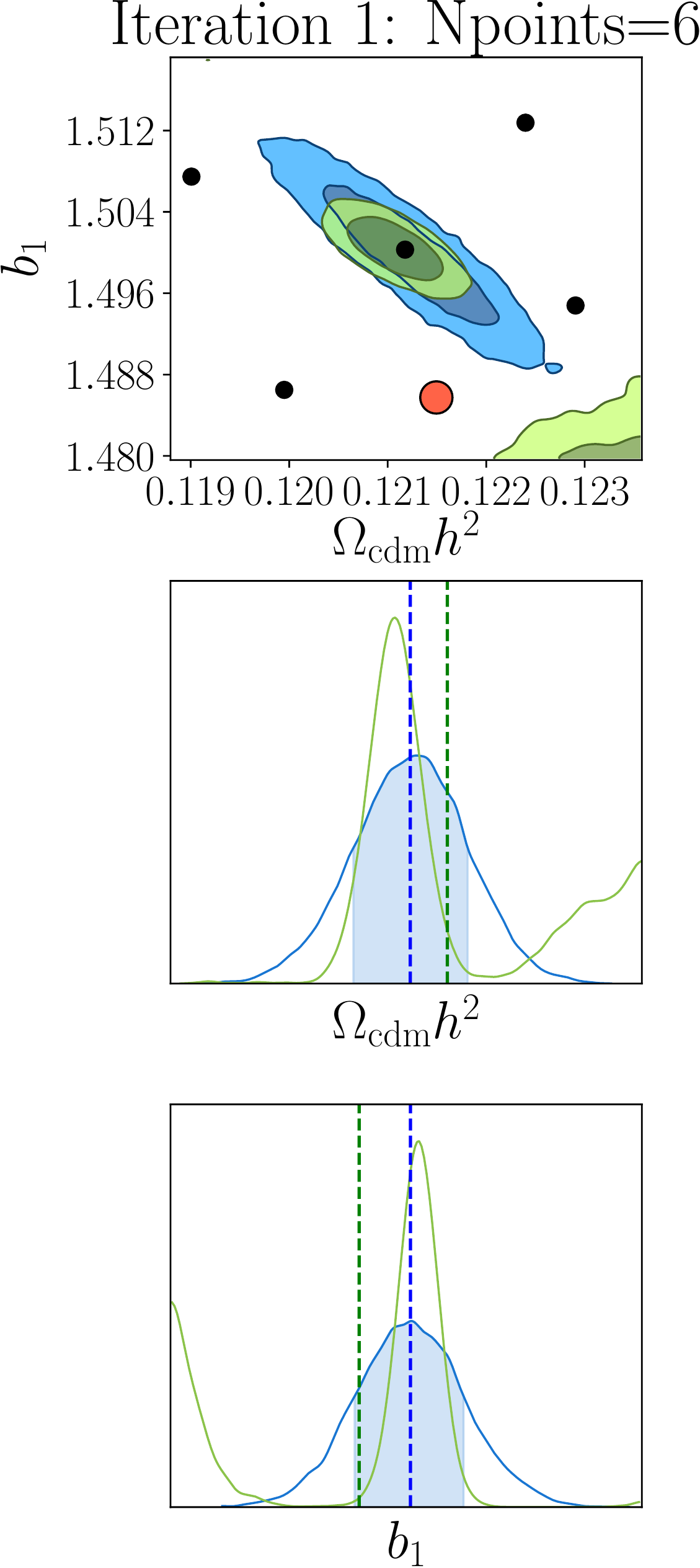}
\includegraphics[width=0.15\textwidth]{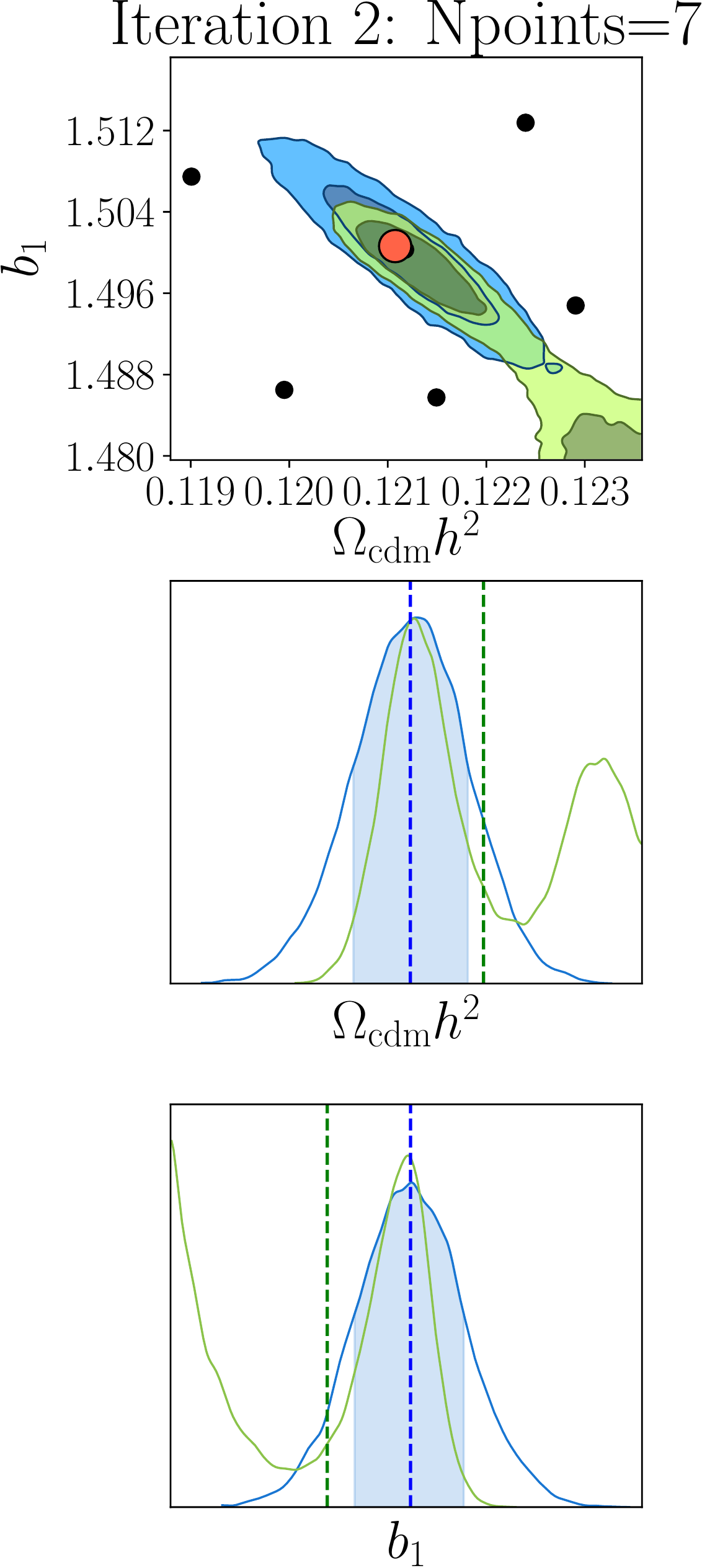}
\includegraphics[width=0.15\textwidth]{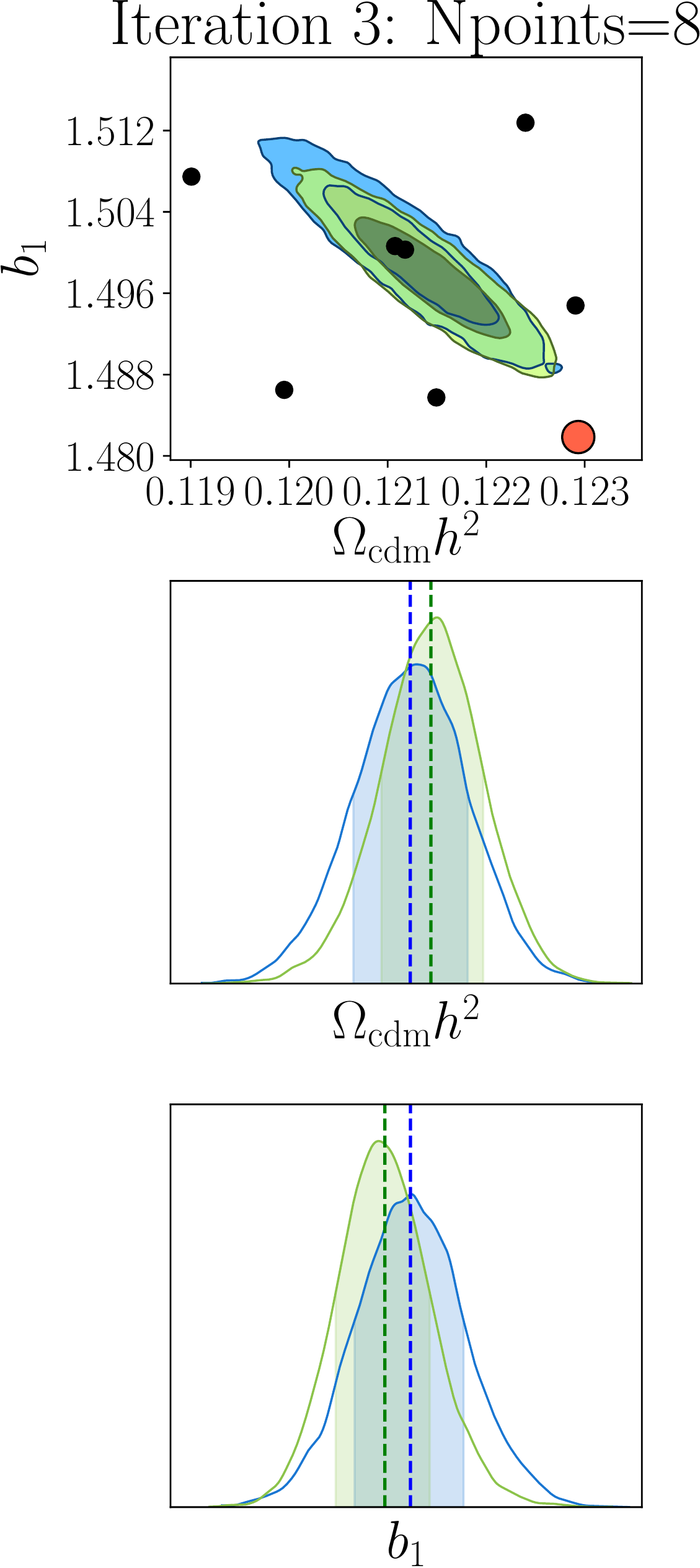}
\includegraphics[width=0.15\textwidth]{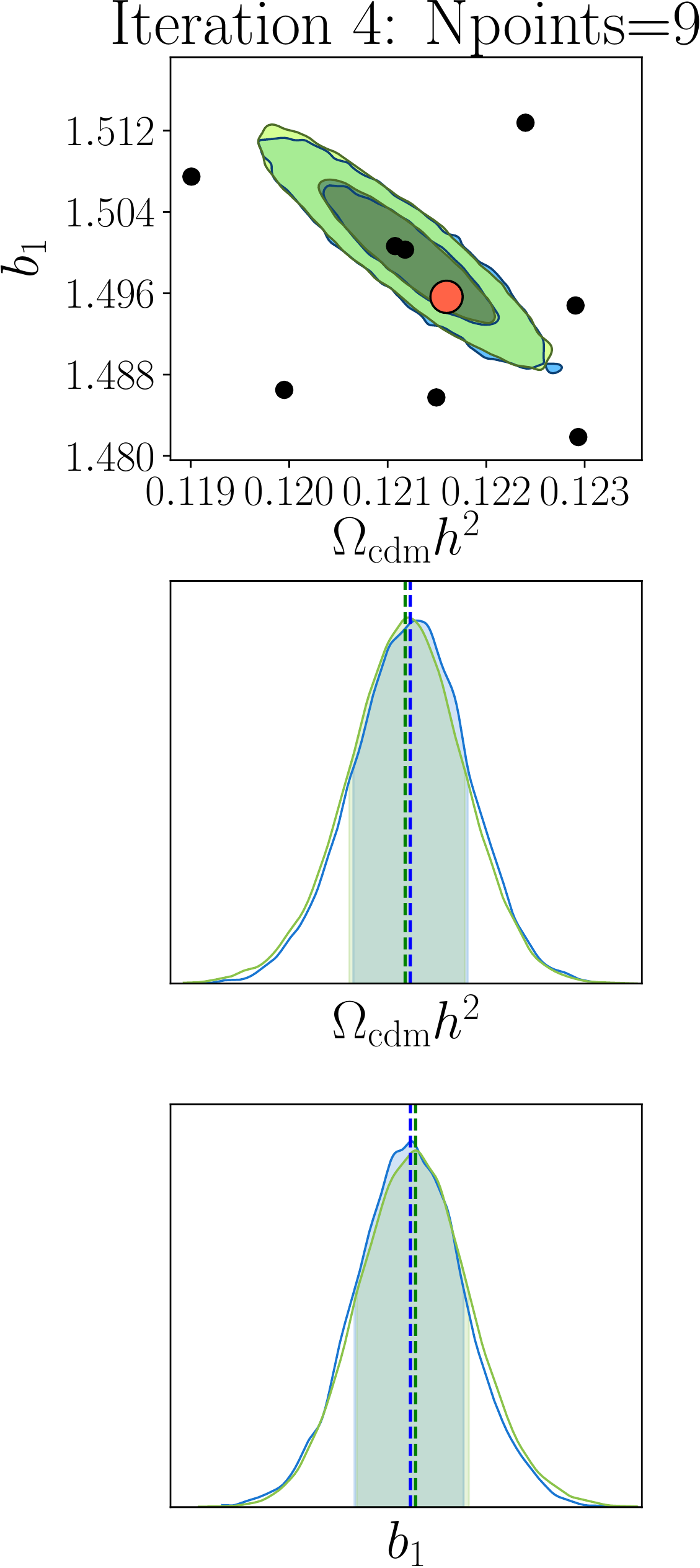}
\includegraphics[width=0.1535\textwidth]{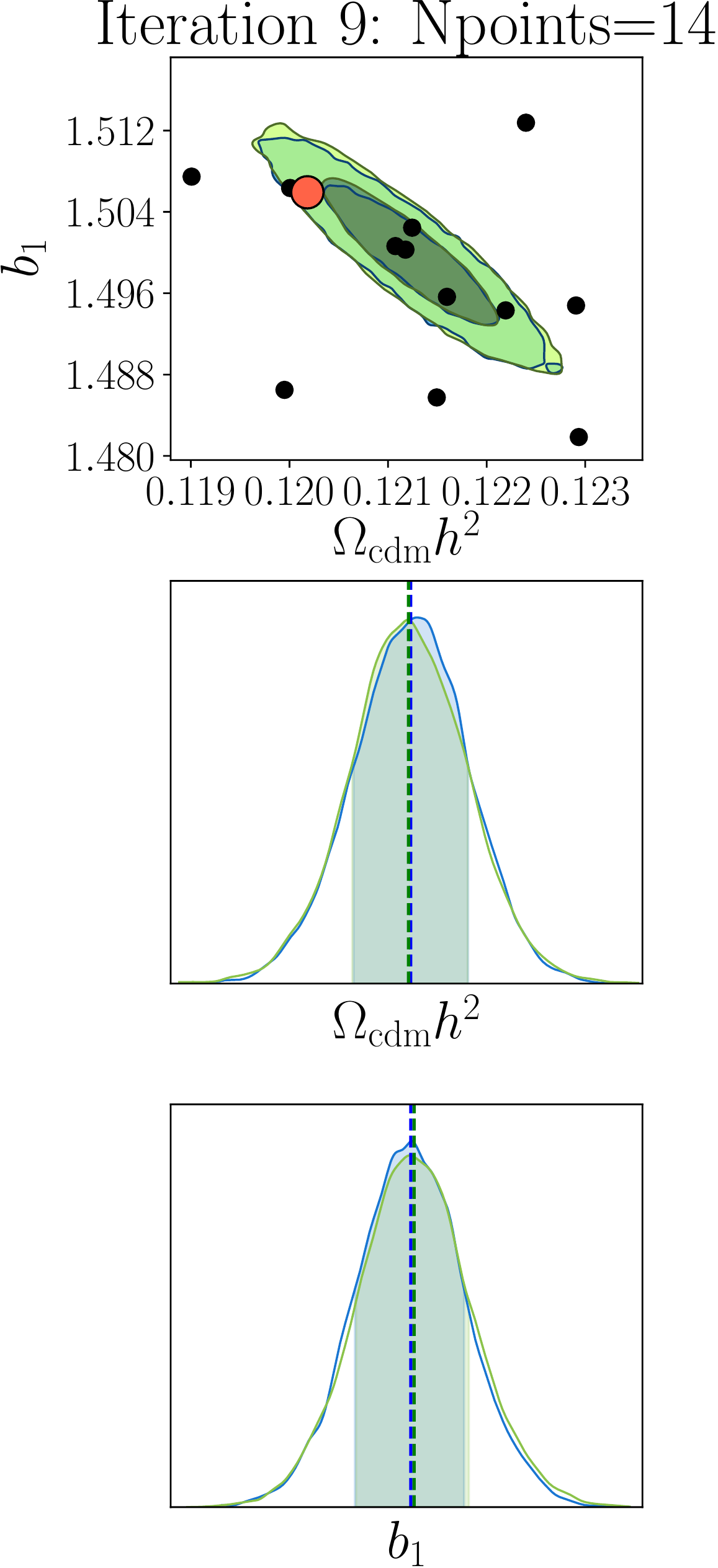}
\caption{Iterative likelihood emulation for an example with two free parameters ($b_1$, $\Omega_{\rm cdm}\,h^2$). The title in each column indicates the iteration number and the corresponding number of model evaluations. The top row displays the two-dimensional posterior on these free parameters, whereas middle and bottom rows display the marginalised posterior distribution functions. The mean of these posteriors is shown by vertical dashed lines. Within each panel, green colour denotes the results of our iterative procedure, whereas blue colours display the results of a standard MCMC with $1000$ steps. In the top panel, black symbols indicate the points where the likelihood has been evaluated by the iterative emulation, and red symbols the new point chosen by sampling a given acquisition function. The displayed ranges coincide with the (flat) prior distribution assumed. }
\label{fig:iterative_procedure}
\end{figure*}

\subsection{Iterative procedure: Bayesian optimization}\label{sec:iter}

In the previous two sections we have discussed how to estimate the probability of $\log \LL(\boldsymbol{\boldsymbol{y}})$ given a theory uncertainty and a training set (likelihood evaluations over a set of parameter values). Naturally, in certain regions of parameter space, the probability will be narrow and the predictions accurate, whereas in other regions (typically far from points where $\LL$ has been evaluated), the uncertainty will be large. 

The basic idea we discuss in this subsection is to adaptively expand the training set by evaluating $\LL$ in regions with high uncertainty but also of high probability contributing to the posterior distribution functions of the parameters of interest \citep[see][for a similar idea applied to the emulation of the Ly-$\alpha$ power spectrum]{Rogers2019}. 

Specifically, we define the following acquisition function
\begin{equation}
 A(\boldsymbol{\theta}) \equiv \LL_{\rm{emul}}+\alpha\,\sigma_{\rm{emul}},
\label{eq:aquisition}
\end{equation}
\noindent where $\mathcal{L}_{\rm{emul}}$ is the emulated likelihood, $\sigma_{\rm{emul}}$ is the standard deviation as estimated by the Gaussian process, and $\alpha$ is a free parameter that balances sampling regions with high uncertainty. Note that, since the Normal distributed parameter is $\log \LL(\boldsymbol{y})$, $\sigma_{\rm{emul}}$ corresponds to the lognormal standard deviation $\sigma_{\rm{emul}}\equiv \left[ e^{\sigma}-1\right]e^{2\chi_{\rm{emul}}^2+\sigma}$. New parameter values for the training set are then sampled with a probability proportional to $A(\boldsymbol{\theta})$. We then recompute $\mathcal{L}_{\rm{emul}}$ given the new set of $\LL$ (with the old evaluations plus the new ones) and iterate this procedure.

To define convergence in our iterative method we make use of the Kullback-Liebler (KL) divergence. The KL quantifies the degree to which two distribution functions, $P$ and $Q$, differ. In the case that the two distributions are multivariate Gaussian with means and covariances ($\boldsymbol{\mu}_q$,$\boldsymbol{\mu}_p$), and ($\Sigma_p$, $\Sigma_q$), it is possible to show that the KL divergence is:

\begin{multline}
 D_{\rm{KL}}(P||Q)= \\ \frac{1}{2} \left[ \log\frac{|\Sigma_q|}{|\Sigma_p|} -d + {\rm{tr}}\left(\Sigma_q^{-1}\Sigma_p\right) + (\boldsymbol{\mu}_q-\boldsymbol{\mu}_p)^{\rm{T}}\Sigma_q^{-1}(\boldsymbol{\mu}_q-\boldsymbol{\mu}_p) \right]   \, ,
 \label{equation:Kullback-Liebler}
\end{multline}

\noindent where $d$ is the number of dimensions of the parameter space, $\rm{tr}$ indicates the trace and $|*|$ refers to the determinant. 

We employ the KL divergence between two consecutive steps in our iterative scheme to estimate how much information newly-added training points provide. If $D_{\rm{KL}} \rightarrow 0$, these newly-added training points do not include any extra information and the process can be considered as converged. We set a threshold value of $D_{\rm{KL}}\approx 0.1$, which roughly correspond to differences in the mean of the multivariate Gaussian posteriors of $5\%$ of their standard deviation.

\subsection{A first example}

Before describing our algorithm in detail, we first consider a simple case that illustrates different aspects and ideas of our approach.

In Fig.~\ref{fig:iterative_procedure} we show the constraints of a two-dimensional parameter space assuming a Gaussian likelihood. The top panels display the joint constraints on these parameters and the middle and bottom panels show the marginalised posterior distribution functions. From left to right, columns show the various steps in our iterative procedure. The color  green denotes the results of our iterative Gaussian emulation, whereas blue indicates results using a standard MCMC algorithm (which is repeated in all panels for comparison). In all panels, black circles indicate the training set and red symbols the value added in a particular iteration.

The leftmost column shows the starting point, where we build the likelihood emulation using only 5 points.  After the first step, we have a rough idea of the function across the entire space, but the large mean distance between sampling points does not permit an accurate emulation, especially if the region of interest is much smaller than the prior hyperspace. 

In the second leftmost panel we sample the emulated function, and select the most convenient points that will be added to the training set. This selection considers both the value and uncertainty of the emulated likelihood since we are not interested in low-likelihood regions or in regions that are already accurately emulated. 

We see that a total of $6$ likelihood evaluations is enough to start roughly identifying the high likelihood region. However, a significant misestimation occurs at the corner of the parameter space. Emulators perform extremely well interpolating but can yield catastrophic extrapolations. This could be particularly worrisome in high dimensional hyperspace due to the large number of corners. However, the iterative process detects these regions of high likelihood and high uncertainty, and samples  them, which consequently improves the emulation. This is in fact what we observe in further iterations.

Finally, after $4$ iterations and $9$ likelihood evaluations, the emulator is converged. The marginalised constraints are to a good degree unchanged as shown in the last two panels of Fig.~\ref{fig:iterative_procedure}), where the contours stay still even though 5 more iterations were performed. Note that these constraints agree with those estimated using an MCMC, displayed in blue, but the latter required approximately $~1000$ evaluations.  

In the next section we will explore in detail this iterative likelihood emulation.

\section{The Algorithm} \label{Sec:algorithm}

In this section we describe in detail our algorithm and the various choices we have made for its implementation.

The first step is to define a small set of points in parameter space from which the emulator can be initially constructed. The least informative way to choose it is for it to cover the whole region defined by the priors. We have adopted a standard Latin-hypercube (\citealt{10.2307/1268522}), modified to maximize projected distances along parameter axes, to define the parameter values. This step was shown in first panel of Fig.~\ref{fig:iterative_procedure} where 5 points were used for the initial construction.

The second step is to define the space and metric to be emulated. Since the kernel assumed for the Gaussian process in Eq.~\ref{eq:kernel} has only one correlation length, we rotate and normalize the space so that it becomes better described by an isotropic kernel. In practice, in each step, the sampling points, $\boldsymbol{s}$, are transformed as

\begin{equation}
s_i \rightarrow (R^s_{ij} \, s_j - m_i)/ \sigma_i
\end{equation} 

\noindent where $\boldsymbol{m}$ and $\boldsymbol{\sigma}$ are the means and standard deviations of the components of the vector $R^s\boldsymbol{s}$ along every dimension, and $R^s$ is a rotation matrix given by $C^s = R^s \Lambda R^s$ where $C^s$ is the covariance matrix of $\boldsymbol{s}$, and $\Lambda$ is a diagonal matrix.

Note that during the first steps, the sampling points are not enough to give an accurate estimation of the rotation matrix $R^s$, but after a certain amount of steps it will converge to the correct rotation matrix together with the emulated likelihood.   

The third step is to build the emulator for $\log \LL$. In this work we will make use of the \texttt{GPy} Python package\footnote{https://sheffieldml.github.io/GPy/} \citep{gpy2014}. Since the optimisation of the kernel is crucial for the correct interpolation of the function, \texttt{GPy} offers an option of restarting the minimisation from different random seeds (\texttt{num\_restarts}) so that one can avoid falling into a local minimum.  

We then fit a multivariate Gaussian to the emulated likelihood and estimate the mean and covariance. We use these quantities to estimate the Kullback-Liebler divergence with respect to the values in the previous iteration. We iterate adding $N_i$ new points to our training set, drawn from the acquisition function $A(\theta)$, regulated through the exploration/exploitation parameter $\alpha$. Thus, the total amount o free parameters of the iterative process are $\{$ $N_i$, $\alpha$, \texttt{num\_restarts}$\}$. If $D_{\rm{KL}} < 0.1$ for $5$ consecutive steps, we determine the process has converged and stop the iterations. Otherwise, we sample $N_i$ new training points where we evaluate the full likelihood and start a new iteration.

In \S\ref{subsec:tests} we will discuss the performance and the impact of various choices for the emulation of galaxy clustering likelihood.


\section{Parameter estimation from the emulation of galaxy clustering likelihood} \label{Sec:sampling}

In this section we present the results of applying the iterative emulation to the estimation of free parameters from mock galaxy clustering data. We start by defining our theoretical description of galaxy clustering (\S\ref{subsec:gal_clust}). We then describe our mock data and respective covariance matrix (\S\ref{subsec:mock_data}). We define the likelihood and provide details of our strategy for parameter estimation (\S\ref{subsec:like}). The results are provided in \S\ref{subsec:results} for cases with and without noise in the data model. We close this section by discussing the impact of various algorithmic choices and the performance of the iterative emulator (\S\ref{subsec:tests}). 

\subsection{Theoretical model for Galaxy clustering}\label{subsec:gal_clust}

We will study the case of galaxy clustering in Fourier space with redshift space distortions (RSD). For this exercise, we assume the anisotropic galaxy power spectrum is given by:

\begin{equation}
 \label{eq:perfect_data}
  P_{\rm g}(k,\mu_k) = b_1^2\left[1+\beta\mu_k^2\right]\left[P_{\rm l}(k) + A k + Q k^2\right]\,e^{-k\sigma_v\mu_k} \, ,
\end{equation}

\noindent where $b$ is a linear bias parameter, $\sigma_v$ is the Gaussian velocity dispersion, and $\beta=f/b$, with $f$ representing the growth factor. $A$ and $Q$ are two empirical parameters describing departures from the linear mass power spectrum, $P_{\rm l}$. Finally, $\mu_k = \hat{k} \cdot \hat{k_z}$ is the cosine angle with respect to the line of sight.

In this particular example we will consider Legendre multipoles of the galaxy power spectrum, defined as:

\begin{equation}
P_\ell(k)=\frac{2\ell+1}{2}\int_{-1}^{1}d\mu_k\,P_g(k,\mu_k)\mathcal{P}_\ell(\mu_k),
\label{eq:pkpoles}
\end{equation}

\noindent where $\mathcal{P}_\ell(\mu_k)$ are Legendre polynomials of order $\ell$. 

This simple model is specified by 9 parameters: 

\begin{equation}
\theta = \{ \Omega_{\rm{cdm}}h^2,\, \Omega_{\rm{b}}h^2,\, h,\, n_s,\, A_s,\, b_1,\, \sigma_v,\, A,\, Q \}\, ;
\end{equation}

\noindent 5 cosmological: the total mass and baryon density in units of the critical density, $\Omega_{\rm{cdm}}$ and $\Omega_{\rm{b}}$; the present-day Hubble constant in units of $100\,{\rm  km s^{-1}\,Mpc^{-1}}$, $h$; the primordial spectral index, $n_s$; the amplitude of the primordial fluctuations, $A_s$; 1 RSD parameters, $\sigma_v$; and $3$ nuisance parameters $b_1$, $A$ and $Q$. 

\subsection{Mock clustering data} \label{subsec:mock_data}

Our mock clustering data will consist of a particular random realisation of our theoretical model for $\tilde{P} \equiv \{ P_{\ell=0}(k), P_{\ell=2}(k)\}$, over the wavelength range $k \in [0.01; 3]\,\ihMpc$. Specifically, we set the model parameters to the following values: $ \Omega_{\rm cdm } = 0.27$, $\Omega_{\rm{b}}=0.05$, $h=0.67$, $n_s=0.96$, $A_s=2.13\times 10^{-9}$, $\sigma_v=3h^{-1}$Mpc, $b=1.5$, $A=1.4$, $Q=4$, and evaluate $P_{\rm l}$ using the linear Boltzmann code \texttt{CLASS} \citep{Blas_2011} at $z=0$. The random realisation will be generated using the appropriate data covariance matrix for a sample of number density $\bar{n} = 5\times 10^{-4}\,(\hMpc)^{-3}$ over a volume $V = 1\,h^{-3}\,Gpc^3$.

We describe the covariance matrix of this mock data, $C_g[\tilde{P}(k),\tilde{P}(k')]$, in linear theory assuming Gaussianity. Explicitly, we assume that $P(k,\mu_k)$ follows a Gaussian distribution which leads to the following relation mode-by-mode covariance (see \citealt{Feldman:1993ky} for the monopole),

\begin{equation}
 \label{eq:ps_cov}
 {\rm{C}}\left[P(k), P(k^\prime)\right] = \frac{2 (2 \pi)^3}{V} \delta_\mathrm{D}(k - k^\prime) \left[ P(k, \mu_k) + \bar n^{-1} \right]^2,
\end{equation}

\noindent where $V$ and $\delta_\mathrm{D}$ stand for the volume of the sample and the Dirac delta respectively. Consequently, the multipole covariance matrix is \citep{Grieb:2015bia} :

\begin{equation}
 \label{eq:ps_ell_cov_bin}
 C_{{\ell_1}{\ell_2}}(k_i,k_j) = \frac{2 (2 \pi)^4}{V_{k_i}^2} \delta_{ij} \int_{k_i-\Delta k/2}^{k_i+\Delta k/2} \sigma^2_{\ell_1\ell_2}(k) k^2 {\rm{d}} k \, ,
\end{equation}

\noindent where the volume of the shell in $k$-space is $V_{k_i} = 4 \pi [(k_i + \Delta k/2)^3 - (k_i - \Delta k/2)^3] / 3$ and 

\begin{multline}
 \label{eq:ps_cov_ell_ell_in_ps_ell}
 \sigma^2_{\ell_1\ell_2}(k) = \frac{2 (2 \ell_1 + 1) (2 \ell_2 + 1)}{V} \\ \times \sum_{\ell_3=0}^\infty \sum_{\ell_4=0}^{\ell_3} \left[ P_{\ell_4}(k) + \frac{1}{\bar n} \delta_{\ell_4 0} \right] \left[ P_{\ell_3 - \ell_4}(k) + \frac{1}{\bar n} \delta_{(\ell_4 - \ell_3) 0} \right] \\
 \times \sum_{\ell=\max(|\ell_1-\ell_2|,|2\ell_4-\ell_3|)}^{\min(\ell_1+\ell_2,\ell_3)} \begin{pmatrix} \ell_1 & \ell_2 & \ell \\ 0 & 0 & 0 \end{pmatrix}^2 \begin{pmatrix} \ell_4 & \ell_3 - \ell_4 & \ell \\ 0 & 0 & 0 \end{pmatrix}^2,
\end{multline}

\noindent where terms in round parenthesis represent Wigner 3j-symbols. 

\begin{figure*}
\centering
\includegraphics[width=0.9\textwidth]{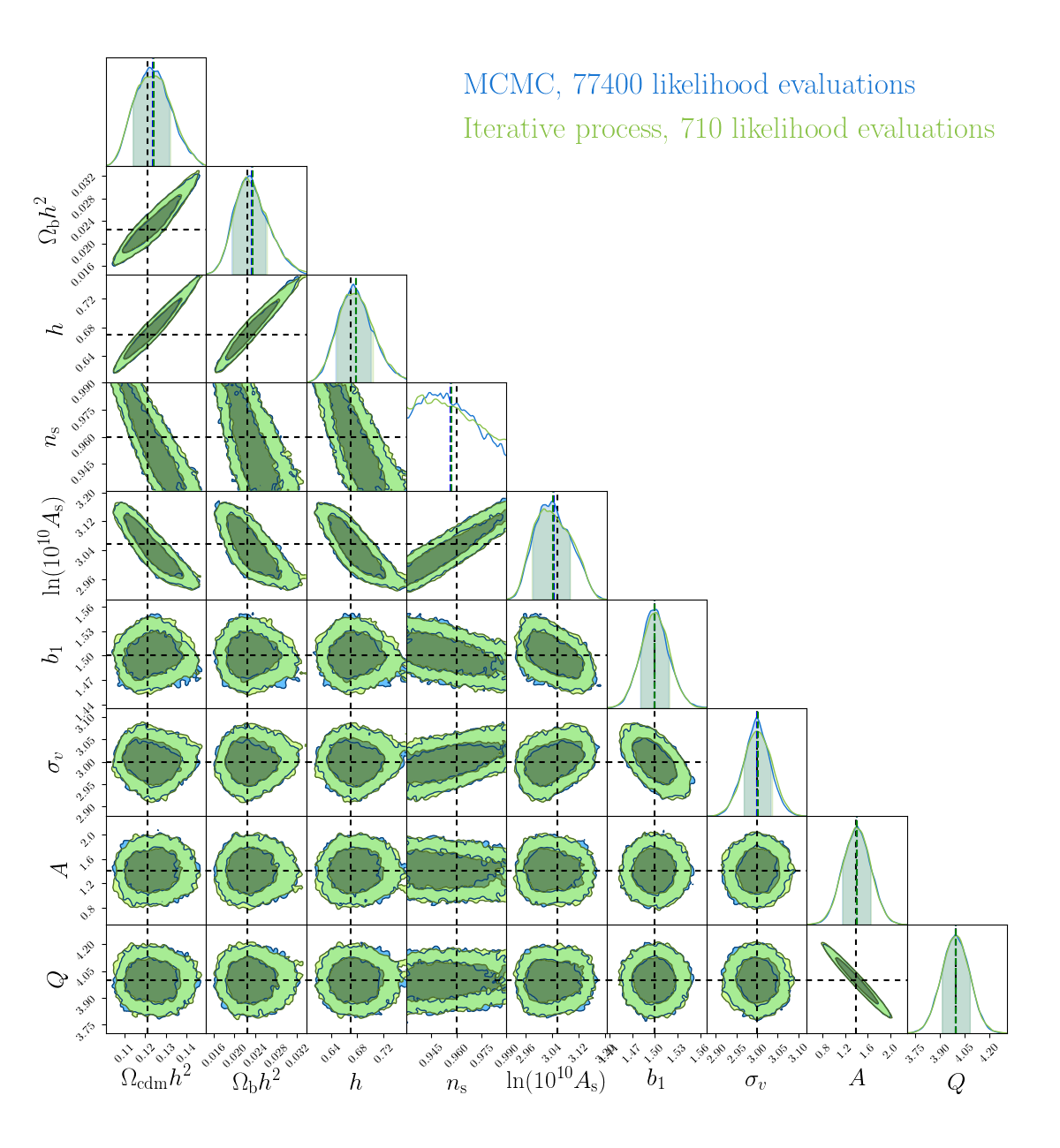}
\caption{Posterior distributions on the $9$ free parameters of our model from the monopole and quadrupole of a mock redshift-space galaxy power spectrum up to $k = 3\,\ihMpc$ in the monopole and $k = 0.3\,\ihMpc$ in the quadrupole. The green and blue contours denote the results from our iterative emulation and a traditional MCMC analysis, respectively. Both methodologies agree to a remarkable level, with the iterative emulator requiring approximately $100$ times less likelihood evaluations. Black dashed lines represent the input values of our mock data. Shaded regions corresponds to the and two $\sigma$ intervals of the projected one dimensional posteriors, i.e. 68$\%$ and 95$\%$ confidence levels.}
\label{fig:contours_ndims9}
\end{figure*}

\begin{figure}
 \centering
 \includegraphics[width=0.47\textwidth]{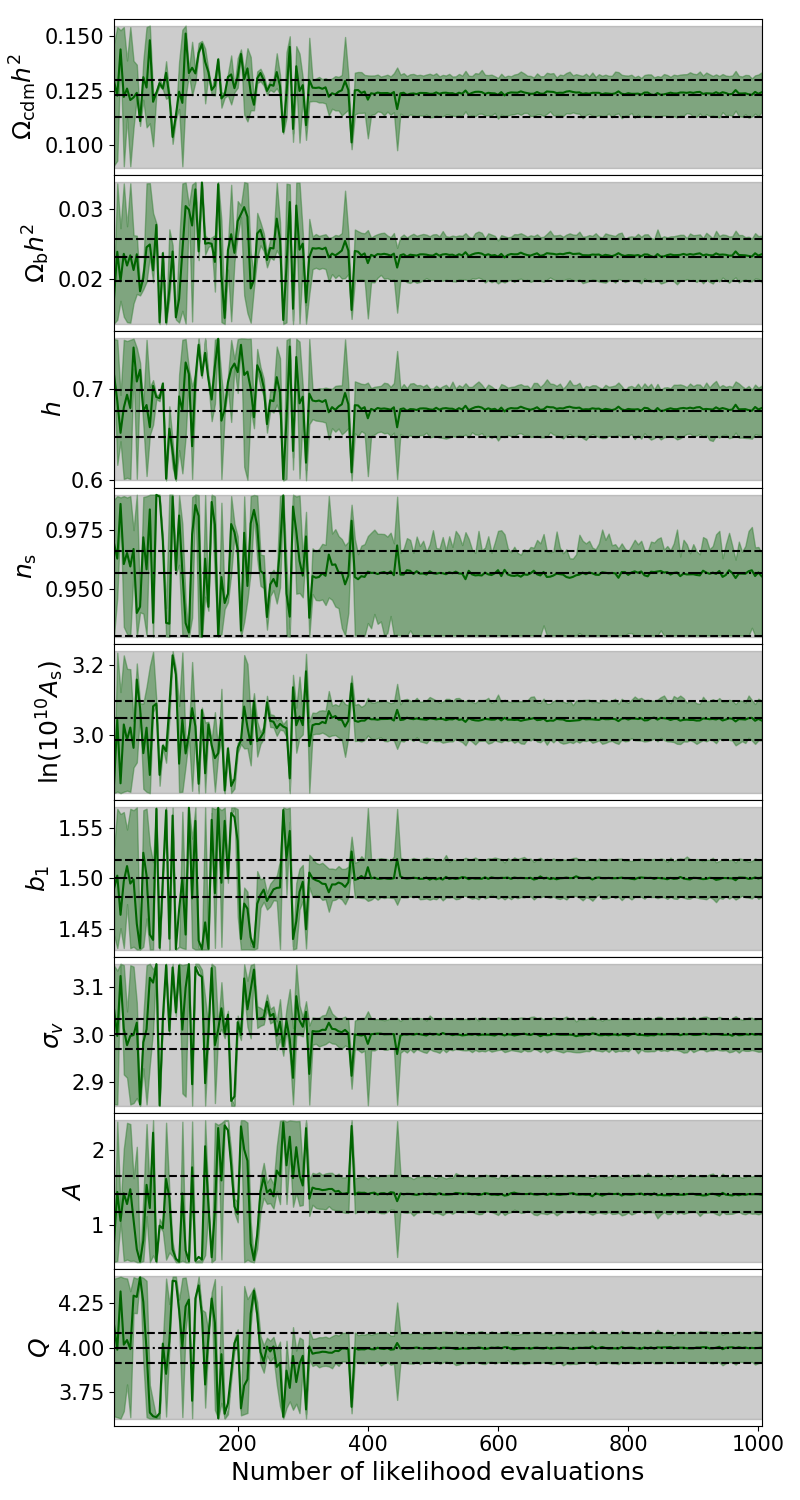}
 \caption{Marginalised constraints for the parameters in our model as a function of the number of likelihood evaluations in our iterative process. Dark green lines show the mean of the posterior distribution, whereas light green shaded regions indicate a region containing 68\% of the distribution. In all cases, the grey region displayed coincides with the ranges of the respective parameter prior distribution. The true value of each parameter with its 68\% confidence level (marked as dashed lines) is indicated as dotted-dashed black lines.}
 \label{fig:mean_values}
\end{figure}

\subsection{Likelihood definition and emulation} \label{subsec:like}

With the assumptions described before, the probability of a given set of multipole values $\tilde{P_d}$ is given by a multivariate normal distribution 

\begin{multline}
 \log \LL = -\frac{1}{2} \left(\tilde{P}_{\rm d}(k) - \tilde{P_{\rm t}}(k)\right)^{\rm{T}}\,C_g^{-1}\,\left(\tilde{P}_{\rm d}(k) - \tilde{P_{\rm t}}(k)\right) \\ - \frac{1}{2}\log|C_{\rm{g}}|- \frac{N}{2}\log(2\pi) \, , 
\label{equation:clust_like}
\end{multline}

\noindent where $\tilde{P}_{\rm t}$ denotes the theoretical models of Eqs.~\ref{eq:perfect_data} and \ref{eq:pkpoles}. Note that, by construction our model is a perfect description of the ensemble mean of the data, and thus we expect unbiased constraints. 

For the 9 free parameters of our model, we assume flat priors over the range $\Omega_{\rm{cdm}} \in [0.2,0.345]$, $\Omega_{\rm{b}} \in [0.03,0.0755]$, $h \in [0.6,0.755]$, $n_s \in [0.93,0.99]$, $\log A_s \in [1.2, 3.25]$, $b_1 \in [1.43,1.57]$, $\sigma_v \in [2.85, 3.15]$, $A \in [0.5, 2.4]$, and $Q \in [3.6, 4.4]$. These values were chosen to be roughly 3 times larger than the expected $68\%$ confidence region for each parameter.

In all cases we will emulate the logarithm of this 9-dimensional likelihood function with the algorithm described throughout this paper. The initial training set is given by a Latin-hypercube of 10 points over the full prior volume. We perform the GP hyper-parameter optimisation with 5 different starting points to avoid local minima. At each point of the iteration, we randomly sample 5 new points according to the acquisition function defined in Eq.~\ref{eq:aquisition} with $\alpha=0$ (maximum exploitation over exploration). We will explore the dependencies on these choices of parameters in \S\ref{subsec:tests}. 

For comparison, we also carry out an affine invariant MCMC sampling of the same likelihood. We employ \texttt{emcee}\footnote{https://github.com/dfm/emcee} \citep{Foreman2013} with 22 walkers, each of which starting at random locations of the prior hyper-volume. We consider (and discard) a burn-in phase of $1/3$ of the steps. To estimate the faithfulness of the chains, we perform a Gelman-Rubin test and assume convergence when it has reached a value of $<0.1$. We note that the Gelman-Rubin test is not formally correct with affine invariant MCMC's, for this reason we checked when the standard deviation of the cumulative mean in each walker was lower than 10$\%$ of the standard deviation of the walker (analogously to the Kullback-Liebler divergence test). We found this criterion was satisfied generally after the Gelman-Rubin one. Therefore, we use the Gelman-Rubin test as conservative criterion. We present and discuss our results in the next subsection.

\subsection{Parameter Constraints}\label{subsec:results}

\subsubsection{Deterministic and unbiased theory prediction}\label{Sec:results_certain}

In Fig.~\ref{fig:contours_ndims9} we show the posterior distributions of all the 9 parameters of our model, as estimated from our mock galaxy clustering data. Green contours show the results of applying the iterative process whereas blue colors denote the results of an MCMC analysis.  

Firstly, we highlight that both methods provide remarkable similar results. The marginalised posteriors provided in diagonal panels are almost indistinguishable, whereas 2D contours are compatible within the noise. The agreement is impressive considering that the MCMC chains required $77400$ likelihood evaluations whereas the iterative procedure proposed in this work used only $\sim 700$ likelihood evaluations. We highlight that even in cases with strong parameter degeneracies (e.g. $A$ and $Q$, or $h$ and $h^2 \Omega_{cdm}$), or where the posterior is truncated by the prior distribution (e.g $n_s$), our emulator is able to correctly recover the posterior. Note also that no phase of burn-in was needed in the iterative case, unlike the MCMC analysis, where $1/3$ of the points were taken away. All the previous evaluations are used for the emulation of a given iteration.  

The mean of the marginalised parameters constraints are shown in Fig.~\ref{fig:mean_values} as a function of the iteration in our likelihood emulations. Each panel displays the marginalised mean value as a dark green line and the $68 \%$ credibility region is denoted as a shaded green region. In each panel, the displayed parameter range coincides with the limits of our prior distribution for the respective parameter. We can see that initially, mean values oscillate wildly within the prior volume. However, as the iteration progresses values converge rather quickly, and once they do, they remain converged and rarely fluctuate around the true value (indicated by horizontal dashed lines) by more than $0.1\sigma_{\theta}$. This also tells that, although in Fig.~\ref{fig:contours_ndims9} we showed the iterative emulation at 710 likelihood evaluations, the correct results were already achieved with only $~485$ sampling points.

\subsubsection{Stochastic theory predictions}

Now we move to the case in which the theory is extracted with some noise, i.e. the theory we have access to is nothing but a sample of a larger space of theories. An example of this are the predictions of $N$-body simulations. If one were to compare data to an $N$-body prediction, even in the best case scenario in which the resolution and population models are perfect, one has to take into account that the specific simulation is subject to, at least, cosmic variance uncertainties. We show here that our methodology can deal with such theory models as well.

\begin{figure*}
\centering
\includegraphics[width=0.9\textwidth]{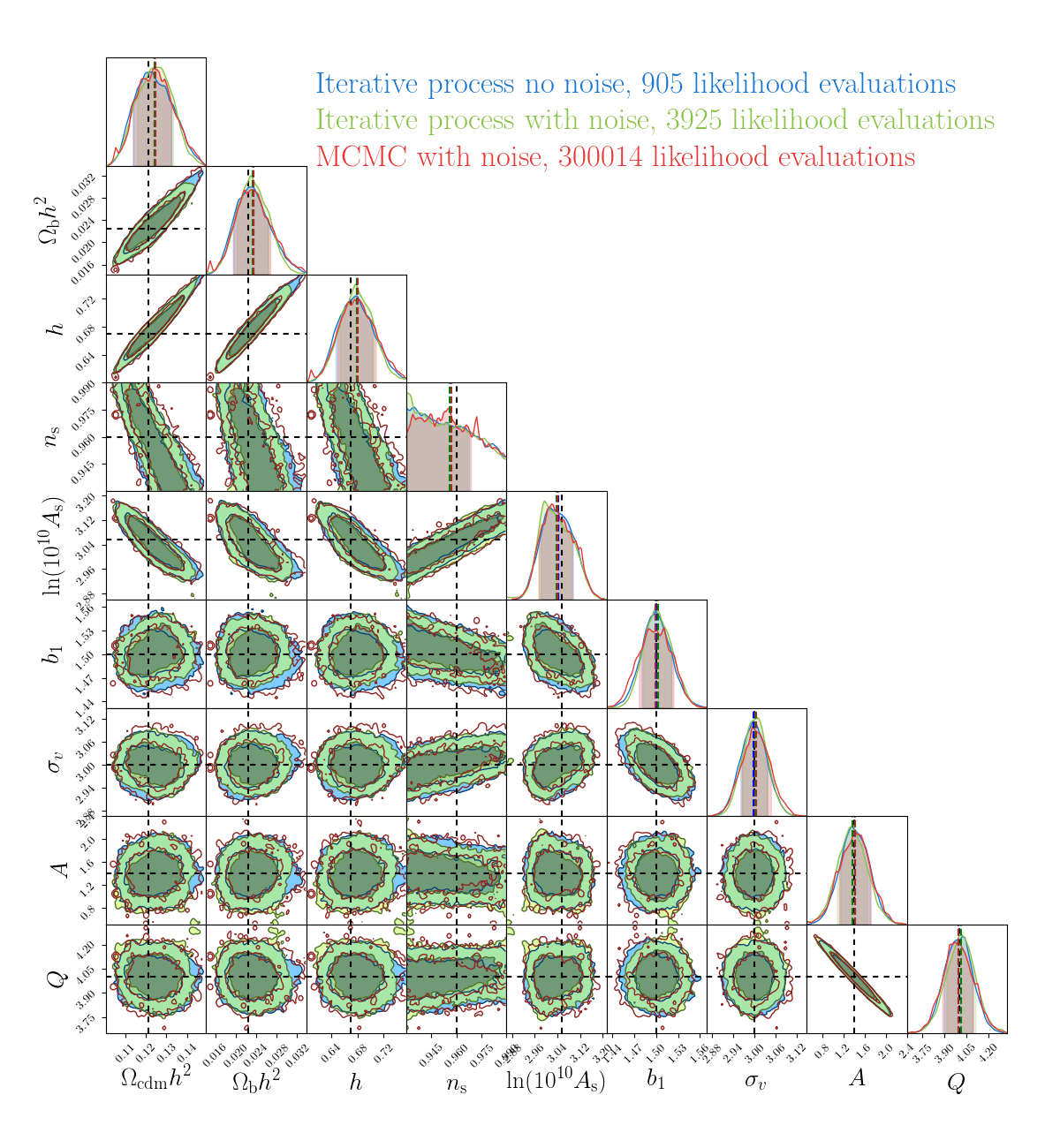}
\caption{Same as Fig. \ref{fig:contours_ndims9} but including a stochastic noise term in the theory modelling and a covariance matrix modified accordingly. Blue colours show the results of the iterative procedure on the deterministic theory model, green contours show the iterative procedure for a noisy theory and brown contours show the MCMC for the noisy theory.}
\label{fig:iterative_procedure_perfect}
\end{figure*}

As a simplification, the noise in the data is assumed to arise only from cosmic variance, thus, it is completely determined by Eq.~\ref{eq:ps_ell_cov_bin}. We further assume that data is only biased with respect to the true cosmology following Eq. \ref{eq:perfect_data}. Therefore, splitting the contributions of the clustering covariance matrix into $C_{\rm{clust}}=C_{\rm{data}}+C_{\rm{theory}}$, our explicit assumption is that $C_{\rm{data}}=C_{{\ell_1}{\ell_2}}(k_i,k_j)$ where the volume and number density ($\bar{n}$) correspond to that of a specific survey. In our case, we chose $\bar{n}=5\times 10^{-4}$ and $V=1000 {\rm{Mpc}}/h$. In general, the noise in the theory is a bit more involved but for the sake of simplicity, we employ Eq.~\ref{eq:ps_ell_cov_bin} once again and assume that the noise in the theory comes from $1/2C_{\rm{data}}$. This resembles the case in which we have perfectly smoothed and unbiased population models and we only suffer from cosmic variance in a simulation with twice the volume as the one from the data survey.       

Applying the iterative process over such kind of theories we find the results of Fig.~\ref{fig:iterative_procedure_perfect}. We show in blue the iterative process without noise in the theory but with the covariance given by the sum of both, theory plus data. We take this to be the true answer to mimic. The iterative process (green contours) is able to recover the previous likelihood well within the 68$\%$ contours even with a noisy theory. Note that, although the noise in the theory translates into a noisy likelihood, the emulator produces smooth results. This is achieved by first sampling the parameter space with 10 Latin-Hypercube points and then iterate 523 times with the addition of 5 points each, reaching a total number of points of 2625. For comparison, we include the converged MCMC in brown. Since the theory is noisy, the MCMC required $~300000$ evaluations of the likelihood to achieve convergence, a factor of $~4$ more points than the non-noisy estimation. This factor is similar in both procedures at this level of noise, which means the iterative process still takes 1/100 of the $\mathcal{L}$ evaluations required by the MCMC. However, we find that the estimation of the MCMC remains noisy even though convergence was achieved.

\subsection{Performace and parameter choices} \label{subsec:tests}

\begin{figure*}
\centering
\includegraphics[scale=0.5]{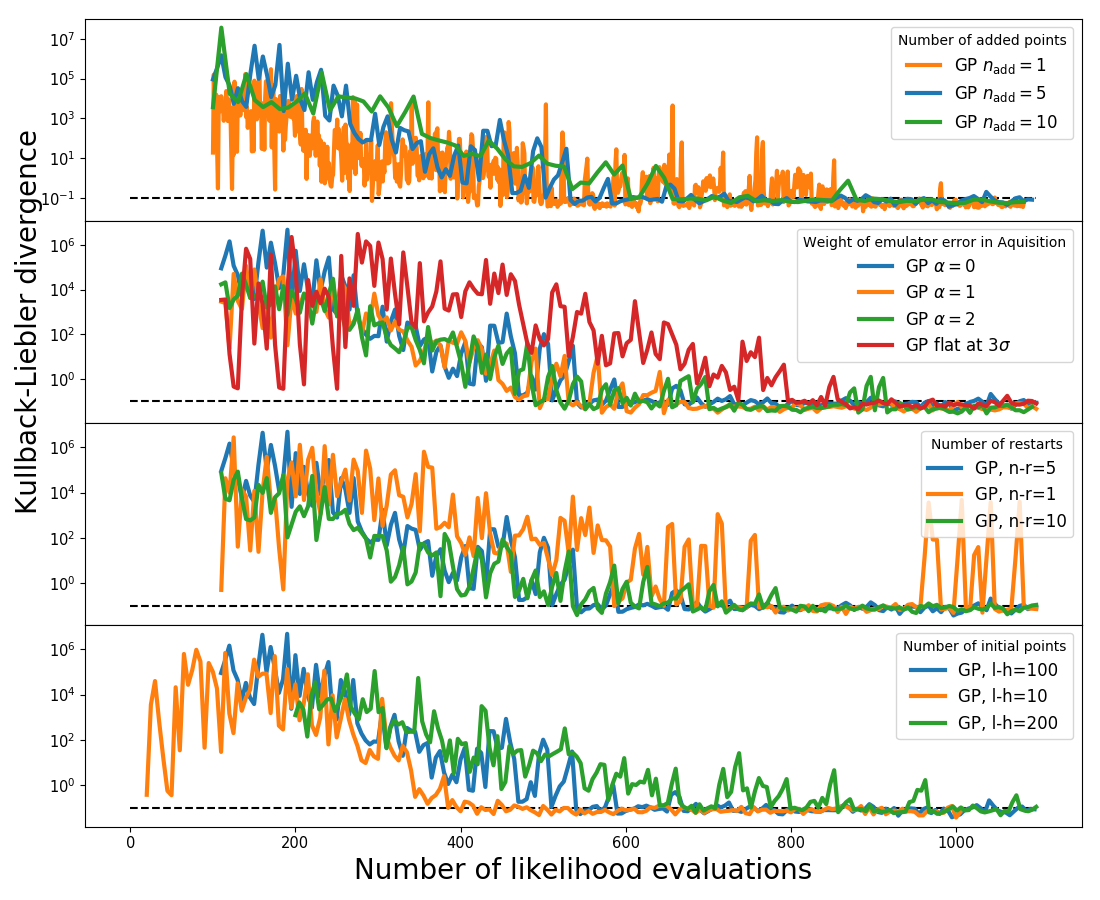}
\caption{Convergence criteria of the iterative process over the parameter space shown in Fig. \ref{fig:contours_ndims9}. GP stands for Gaussian Process, n-r for number of restarts and l-h for the number of Latin-hypercube initial points. Black dashed line represents the value of the Kullback-Liebler divergence $0.1$.}
\label{fig:convergence}
\end{figure*}

In this subsection we will explore how the performance of the iterative likelihood emulator depends on various parameters of the algorithm, as well as its dependence on the dimensionality of the likelihood. 

In Fig.~\ref{fig:convergence} we show the Kullback-Liebler convergence criteria, $D_{KL}$, as a function of the number of likelihood evaluations in our iterative emulations applied to the 9-dimensional galaxy clustering likelihood considered before. Different lines show the results for difference cases as indicated by the legend.

Firstly, we note that regardless of these choices, we observe a clear converging behaviour where $D_{KL}$ decreases exponentially with iteration number. When it reaches a value of $0.1$ it saturates and it coincides with the converged posterior distributions we observed before. This behaviour further motivates the use of a $0.1$ threshold as a convergence criterion.

In the top panel of Fig.~\ref{fig:convergence} we vary $n_{\rm{add}}$, the number of new training points we add in each iteration. As expected, faster convergence is achieved for lower values of $n_{\rm{add}}$ -- $n_{\rm{add}}=1$ requires approximately 200 less likelihood evaluations than  $n_{\rm{add}} = 10$. However, larger values of $n_{\rm{add}}$ converge with less builds of the emulator and also allow for straightforward parallelisation. Thus, $n_{\rm{add}} > 1$ might be preferred depending on the details of the problem at hand. 

In the second panel we vary the function we employ for acquiring new training points. First, we vary the value of the $\alpha$ parameter from $0$ to $2$, which transitions from pure ``exploitation" -- sampling preferentially high-likelihood regions, to ``exploration" where the focus is on highly uncertain regions (c.f. Eq.~\ref{eq:aquisition}). We see that the behaviour is fairly insensitive to this value in terms on how many evaluations are needed to achieve convergence, which indicates that the likelihood itself largely dominates the acquisition function. We have also experimented with an alternative definition of acquisition function in which the likelihood value is capped to a fixed value within a $3\sigma$ region around the likelihood peak -- so that all points within this region have the same probability of being chosen. This alternative acquisition function is displayed as a red line, but we see that it underperforms compared to our default choice.

In the third panel of Fig.~\ref{fig:convergence} we can see how the convergence is affected by the number of restarts in the emulation hyper-parameter fitting. Smaller values lead to faster code execution, however, this degrades the convergence by up to $200$ evaluations as an inaccurate emulation will propagate by sampling suboptimally new training points. Furthermore, it seems that the choice of \texttt{num\_restarts}=5 already saturates and provides optimal results.

In the bottom panel of Fig.~\ref{fig:convergence} we see that the number of evaluations in the initial training set does not strongly impact the long-term behaviour of the emulated likelihood. On the contrary, smaller initial training sets require a smaller number of total evaluations for convergence. This supports the overall performance of our algorithm, which locates new likelihood evaluations optimally.

Finally, we consider the performance of our iterative emulation as a function of the dimensionality of the problem. For this, we have considered a subset of our full parameter space with different number of dimensions, from 2 to 9. In each case, we sample the respective likelihood with an MCMC algorithm and with our iterative emulator. In Table~\ref{table:mcmc} we provide the number of likelihood evaluations required to achieve convergence for different number of dimensions for both {\tt emcee} and our iterative emulator. Respectively, as for our fiducial case, we define convergence with a Gelman-Rubin criterion with a threshold of $0.1$, or with a KL divergence threshold of $0.1$.

\begin{table}
\centering
\begin{tabular}{ |c|c|c|c|c| } 
\hline
$n_{\rm{dims}}$ & MCMC & Iterative & \% points \\
\hline
2 & 8822 & 79 & 0.9\% \\ 
3 & 4862 & 51 & 1.0 \% \\ 
4 & 11044 & 138 & 1.3\% \\
5 & 39622 & 138 & 0.8 \% \\
6 & 33066 & 275 & 0.8 \% \\
7 & 41822 & 435 & 1\% \\
9 & 77400 & 485 & 0.6 \% \\
\hline
\end{tabular}
\caption{Number of likelihood evaluations needed to reach convergence in a MCMC sampler and in our iterative emulation algorithm, for different number of dimensions.}
\label{table:mcmc}
\end{table}

We can see that both likelihood estimators scale roughly linearly with the dimensionality of the problem. The scaling, however, is not perfect since it also depends on the complexity of the target likelihood function. Nevertheless, the relative number of steps is roughly independent of dimensions with the iterative emulator converging with $\sim1\%$ of the evaluations needed by the MCMC sampler.


\section{Summary and Conclusions} \label{Sec:Conclusions}

We have proposed and tested an algorithm for cosmological parameter space sampling. The approach is based on an iterative Gaussian emulation of likelihoods, where new training points are added in regions of high likelihood and large uncertainty, and convergence is determined by the Kullback-Leibler divergence. 

In Fig.~\ref{fig:iterative_procedure} we illustrated our algorithm in action. We then employed it on the problem of constraining 9 cosmological and nuisance parameters using mock data for the monopole and quadrupole of the galaxy power spectrum. In Fig.~\ref{fig:contours_ndims9} and \ref{fig:iterative_procedure_perfect} we showed that our iterative emulation and a standard MCMC agree to a remarkable level, but our approach requires approximately 100 times less model evaluations. In Fig.~\ref{fig:convergence} we showed that this behaviour is roughly independent from the parameters of the algorithm and from the dimensionality of the problem considered.

In general, we found that there are two main advantages of our iterative method over standard MCMC algorithm:

\begin{itemize} 
\item[i)] The use of a small number of likelihood evaluations. This is especially important when using sophisticated models for large-scale structure which become increasingly more computationally expensive. 
\item[ii)] Uncertainties in data models can be naturally incorporated in the likelihood estimation. 
\end{itemize}

In this work we have explored a high-dimensional but relatively simple likelihood function. Although such simple, Gaussian, and unimodal likelihoods are typical in LSS and cosmological analysis, more complicated ones can also appear in
certain situations (e.g. ``banana'' shapes in the the posterior distribution, \citealt{Abbott:2017wau}). To test our algorithm in those situations, we iteratively emulated a non-convex ``Rosenbrock'' likelihood and multi-peaked likelihoods in two dimensions. Although not shown here, we find good performance and an accurate emulation. Nevertheless, for extremely complicated likelihood functions, the assumption of isotropic kernels in our Gaussian emulation might prove limiting. In such cases, anisotropic kernels with multiple hyper-parameters and/or Deep Gaussian processes could be an interesting area of further exploration.

Despite the high efficiency of our approach, there are several areas where it  can be further improved. Although typically building a Gaussian emulation is fast, it scales poorly with the number of sampling points (naively as $\mathcal{O}(n^3)$). This shortcoming can be overcome by recently-proposed methods for matrix inversions \citep{Ambikasaran:2015}. Another alternative is to reduce the number of sampling points by determining a minimal set of inducing points to be used in the emulation \citep{Titsias:2009,Matthews:2015}. Finally, using information on the hyper-parameter minima of previous iterations, might speed up the algorithm overall. 

We anticipate our approach will enable more accurate modelling of cosmological observables and LSS in particular. An interesting application for the latter is the use of cosmology-rescaling methods \citep{Angulo:2010,Zennaro:2019aoi}, where one gravity-only simulation is used to represent the 3-dimensional matter and baryonic fields \citep{Arico:2019}. Although these methods are very accurate and fast compared to a full simulation (Contreras et al in prep), they are very slow for a traditional MCMC sampler. Our work might enable their use in cosmological analyses, which should in general yield to a more complete analysis of current and future observations.


\section*{Acknowledgements}

We acknowledge useful discussions with Eduardo Rozo and Carlos Hernandez-Monteagudo. The authors acknowledge the support of the ERC Starting-Grant 716151 (BACCO). SC acknowledges the support of the ``Juan de la Cierva Formaci\'on'' fellowship (FJCI-2017-33816).

\textit{Author contributions:} MPI, REA and GA provided the main theoretical background and original ideas of this work. GA and MPI developed the core of the algorithm and the coding. MZ, SC and JS tested ideas, helped with the coding and further developed different parts of the methodology.




\bibliographystyle{mnras}
\bibliography{bibliography} 


\bsp	
\label{lastpage}
\end{document}